\documentclass[secnumarabic, prb, showkeys]{revtex4}
\usepackage{bm}

\newcommand{\s}{\circle*{0}}

\begin{document}

\title{Mechanical scenario for the reaction:\\
neutron $\to$ proton  $+$ electron $+$ antineutrino}
\author{Valery P. Dmitriyev}
\affiliation{Lomonosov University\\
P.O.Box 160, Moscow 117574, Russia}
\email{aether@yandex.ru}
\date{1 February 2006}

\begin{abstract}
Small perturbations of averaged ideal turbulence reproduce the
electromagnetic field. The luminiferous medium has the relatively
low pressure and high energy. A vapor bubble in the fluid models
the neutron. The bubble stabilized via creating in the fluid a
field of the positive perturbation of the turbulence energy serves
as a model of the proton. An isle of the fluid that produces the
respective field of the negative perturbation of the turbulence
energy models the electron. The antineutrino corresponds to a
local positive disturbance of the turbulence energy needed in
order to compensate the difference in perturbations of the energy
produced by the electron and proton.
\end{abstract}
\maketitle \maketitle

\section{Substratum for physics}

The properties of physical space known as fields and particles can
be modelled in terms of continuum mechanics \cite{Dmitr1, Dmitr2}.
The concept of substratum for physics is used to this end. The
substratum serves as a medium to carry over the electromagnetic
wave and to transmit interactions. On a large scale the substratum
can be approximated by a turbulent medium.

\section{Perturbations of turbulence}

We consider the averaged turbulence of an ideal fluid. The system
is governed by linearized Reynolds equations \cite{Troshkin,
Troshkin1, Dmitriyev}:
\begin{eqnarray}
\partial _i\! \left\langle u_i\right\rangle &=& 0, \label{1}\\
 \varsigma \partial _t\!\left\langle {u_i } \right\rangle  + \varsigma \partial _k\!
 \left\langle {u'_i u'_k } \right\rangle  + \partial _i\!\left\langle p\,
 \right\rangle &=& 0 \label{2}
\end{eqnarray}
where $\varsigma$ is the medium density, $\langle
{\bf u}\rangle$ the averaged velocity of the fluid flow, $\langle
p\,\rangle$ the averaged pressure and ${\bf u'}$ the turbulent
fluctuation of the velocity, so that
\begin{equation}
{\bf u} = \langle {\bf u}\rangle  + {\bf u'}. \label{3}
\end{equation} Here
and further on we use denotations $\partial_t=\partial/\partial
t$, $\partial_k =
\partial/\partial x_k$. Summation over recurrent index is implied
throughout.

We assume that in the unperturbed state the turbulence is
homogeneous and isotropic, i.e.
\begin{eqnarray}
\left\langle {\bf u} \right\rangle ^{\left( 0 \right)} &=& 0,
\label{4}\\
 \left\langle p\, \right\rangle ^{\left( 0 \right)} &=&
p\,_0, \label{5}\\
\left\langle {u'_i u'_k } \right\rangle ^{\left( 0 \right)} &=&
c^2 {\rm \delta }_{ik} \label{6}
\end{eqnarray}
where $p\,_0, c = {\rm const}$ and $c$ is the speed of the
turbulence perturbation wave in the medium. Integrating equation
(\ref{2}) for an isotropic turbulence
\begin{equation}
\langle u'_i u'_k \rangle = \langle u'_1 u'_1 \rangle {\rm \delta
}_{ik}\label{7}
\end{equation}
and $\langle{\bf u}\rangle=0$, we get a kind of Bernoulli equation
\begin{equation}
\varsigma\! \left\langle {u'_1 u'_1 } \right\rangle  +
\left\langle p\, \right\rangle = \varsigma c^2 + p\,_0. \label{8}
\end{equation}
Formally, by (\ref{8}), any distribution of the turbulence energy
density
\begin{equation}
\frac{1}{2}\varsigma\langle u'_i u'_i \label{9}\rangle
\end{equation}
may occur.

The next linearized term in the chain of Reynolds equations looks
as follows
\begin{equation}
\partial _t \langle u'_i u'_k \rangle  + c^2 \left( {\partial _i  \langle u_k \rangle  +
\partial _k  \langle u_i \rangle } \right) + h_{ik}  = 0 \label{10}
\end{equation}
where
\begin{equation}
\varsigma h_{ik}  = \left\langle {u'_i
\partial _k p'} \right\rangle  + \left\langle {u'_k \partial _i
p'} \right\rangle + \varsigma \partial _j \left\langle {u'_i u'_j
u'_k } \right\rangle.\label{11}
\end{equation}
The density (\ref{9}) of the turbulence energy is in a way similar
to distribution of the heat energy. However, the aether fluid is a
true continuum, i.e. it does not consist of corpuscles and hence
does not imply the dissipation of the mechanical energy. It can be
shown\cite{Troshkin1} that in the absence of diffusion
\begin{equation}
h_{ii} = 0. \label{12}
\end{equation}
Taking in (\ref{10}) $i=k$  and summing over the recurrent index
we get with the account of (\ref{1}) and (\ref{12})
\begin{equation}
\partial _t \langle u'_i u'_k \rangle = 0. \label{13}
\end{equation}
By (\ref{13}) the profile of linear perturbations is conserved in
the nondissipative incompressible medium.

For uniformity with (\ref{2}), it is expedient to differentiate
equation (\ref{10}) with respect to $x_k$:
\begin{equation}
\partial _t \partial _k\langle u'_i u'_k \rangle  + c^2 \nabla^2
 \langle u_i \rangle  + \partial _k h_{ik}  = 0 \label{14}
\end{equation}
where the incompressibility condition (\ref{1}) was used.

\section{Maxwell's equations}

With the definitions
\begin{eqnarray}
 A_i  &=& \kappa c\left( {\left\langle {u_i }
\right\rangle  - \left\langle {u_i } \right\rangle ^{\left( 0
\right)} } \right),\label{15}\\
 \varsigma \varphi  &=& \kappa \left(
{\left\langle p\, \right\rangle - \left\langle p\, \right\rangle
^{\left( 0 \right)} } \right), \label{16}\\
 E_i  &=& \kappa \partial
_k \left( {\left\langle {u'_i u'_k } \right\rangle  - \left\langle
{u'_i u'_k } \right\rangle ^{\left( 0 \right)} }
\right),\label{17}\\
 j_i  &=& \frac{\kappa }{{4\pi}}\partial _k h_{ik}\label{18}
\end{eqnarray}
where $\kappa$ is an arbitrary constant, (\ref{2}), (\ref{14}) and
(\ref{1}) take the form of Maxwell's equations
\begin{eqnarray}
\frac{1}{c}\,\partial _t {\bf A} + {\bf E}  + \bm{\nabla} \varphi
&=&
0,\label{19}\\
\frac{1}{c}\frac{{\partial {\bf E}}}{{\partial t}} - \bm{\nabla}
\times \left( {\bm{\nabla}  \times {\bf A}} \right) + \frac{{4\pi
}}{c}{\bf j} &=& 0\label{20}
\end{eqnarray}
with the Coulomb gauge
\begin{equation}
\bm{\nabla} \cdot {\bf A} = 0 \label{21}
\end{equation}
respectively, where in (\ref{20}) the general vector relation
\begin{equation}
\nabla^2 {\bf A} =  \bm{\nabla} (\bm{\nabla} \cdot {\bf A}) -
\bm{\nabla} \times (\bm{\nabla}  \times {\bf A}) \label{22}
\end{equation}
and incompressibility condition (\ref{21}) were used.

 So, small perturbations of an ideal
turbulence reproduce the electromagnetic field \cite{Troshkin}.

\section{The neutron as a vapor bubble}

Discontinuities, or defects, of the medium model particles.
Typically, a spherical cavity included into the medium represents
a point-like discontinuity. A bubble filled with the vapor of the
fluid can be taken as a model of the neutron.

Let $V^*$ be the volume of the turbulent fluid evaporated into the
bubble. The kinetic energy $K^*$ transferred with the fluid into
the gas phase can be found from (\ref{9}). We get with the account
of (\ref{6}) for the unperturbed medium
\begin{equation}
K^* =  V^*\frac{3}{2}\varrho\langle u_1'u_1'\rangle^{(0)} =
\frac{3}{2}\varrho V^*c^2.\label{23}
\end{equation}
The vapor will be assumed to behave as an ideal gas. The equation
of state of the ideal gas can be written in the form
\begin{equation}
pV = \frac{2}{3}K.\label{24}
\end{equation}
In the mechanical equilibrium the gas pressure $p$ must be equal
to the fluid pressure $\langle p \rangle$. If $V$ is the volume of
the bubble then using (\ref{23}) in (\ref{24}) we get for a bubble
in the unperturbed medium
\begin{equation}
p_0V = \varrho V^*c^2 \label{25}
\end{equation}
where $p_0$ is the background pressure (\ref{5}). Insofar as
$V^*\ll V$ we see from (\ref{25}) that the occurrence in the fluid
of ideal vapor bubbles implies the turbulence of the high energy
and low pressure:
\begin{equation}
p_0  \ll \varsigma c^2.\label{26}
\end{equation}

\section{The mass and self-energy of a particle}

The mass of the bubble can be determined from the mass of the gas
contained in it
\begin{equation}
m = \varrho V^*.\label{27}
\end{equation}
Substituting (\ref{27}) into (\ref{25}) we get
\begin{equation}
p_0V = mc^2.\label{28}
\end{equation}
The self-energy of a bubble can be defined as the work needed in
order to create the bubble in the unperturbed medium
\begin{equation}
E = p_0V.\label{29}
\end{equation}
Using (\ref{29}) in (\ref{28}) gives
\begin{equation}
E = mc^2.\label{30}
\end{equation}

In the phenomenological theory (see e.g. the
textbook\cite{Irodov}) we can deduce for the increment of the
kinetic energy of a particle
\begin{equation}
dE = c^2dm.\label{31}
\end{equation}
In order to obtain from (\ref{31}) the expression for the
self-energy we must postulate that all the internal energy of the
particle concerned with the mass has the kinetic origin. As you
see from the microscopic theory considered the self-energy
(\ref{30}) of a particle is indeed immediately reduced to a part
of the turbulence energy of the luminiferous medium.

\section{Conversion to the proton}

\begin{figure}[h]
\begin{picture}(160,100)(0,-20)
\put (0,-10){\vector(1,0){140}} \put(0,-10){\vector(0,1){90}}
\multiput(-3,0)(0,10){8}{\line(1,0){3}}
\multiput(30,-13)(30,0){4}{\line(0,1){3}}

\put(-15, 57.5){$\varsigma c^2$} \put(-15, 7.5){$p\,_0$}
\put(90,50){$\varsigma\langle u^\prime_1 u^\prime_1 \rangle $}
\put(105,15){$\langle p\,\rangle$} \put(135,-20){$x$} \put(120,
90){}\put(-10, -13.5){0}


\put(0.0,   64.0){\circle{0.01}}\put(0.805369,   64.0544
){\s}\put(1.61074, 64.1103 ){\s}\put(2.41611, 64.1678
){\s}\put(3.22148, 64.227 ){\s}\put(4.02685, 64.2878
){\s}\put(4.83222, 64.3504 ){\s}\put(5.63758, 64.4148
){\s}\put(6.44295, 64.4812 ){\s}\put(7.24832, 64.5496
){\s}\put(8.05369, 64.6202 ){\s}\put(8.85906, 64.6929
){\s}\put(9.66443, 64.768 ){\s}\put(10.4698, 64.8455
){\s}\put(11.2752, 64.9256 ){\s}\put(12.0805, 65.0084
){\s}\put(12.8859, 65.094 ){\s}\put(13.6913, 65.1826
){\s}\put(14.4966, 65.2743 ){\s}\put(15.302, 65.3694
){\s}\put(16.1074, 65.4679 ){\s}\put(16.9128, 65.5701
){\s}\put(17.7181, 65.6762 ){\s}\put(18.5235, 65.7864
){\s}\put(19.3289, 65.901 ){\s}\put(20.1342, 66.0202
){\s}\put(20.9396, 66.1443 ){\s}\put(21.745, 66.2737
){\s}\put(22.5503, 66.4086 ){\s}\put(23.3557, 66.5495
){\s}\put(24.1611, 66.6966 ){\s}\put(24.9664, 66.8506
){\s}\put(25.7718, 67.0118 ){\s}\put(26.5772, 67.1807
){\s}\put(27.3825, 67.358 ){\s}\put(28.1879, 67.5443
){\s}\put(28.9933, 67.7403 ){\s}\put(29.7986, 67.9467
){\s}\put(30.604, 68.1644 ){\s}\put(31.4094, 68.3944
){\s}\put(32.2148, 68.6377 ){\s}\put(33.0201, 68.8955
){\s}\put(33.8255, 69.1692 ){\s}\put(34.6309, 69.4603
){\s}\put(35.4362, 69.7705){\s}



\put(84.5637,    69.7705 ){\s}\put(85.3691,    69.4603
){\s}\put(86.1745,    69.1692 ){\s}\put(86.9798, 68.8955
){\s}\put(87.7852, 68.6377 ){\s}\put(88.5906, 68.3944
){\s}\put(89.3959, 68.1644 ){\s}\put(90.2013, 67.9467
){\s}\put(91.0067, 67.7403 ){\s}\put(91.812, 67.5443
){\s}\put(92.6174, 67.358 ){\s}\put(93.4228, 67.1807
){\s}\put(94.2281, 67.0118 ){\s}\put(95.0335, 66.8506
){\s}\put(95.8389, 66.6966 ){\s}\put(96.6442, 66.5495
){\s}\put(97.4496, 66.4086 ){\s}\put(98.255, 66.2737
){\s}\put(99.0603, 66.1443 ){\s}\put(99.8657, 66.0202
){\s}\put(100.671, 65.901 ){\s}\put(101.476, 65.7864
){\s}\put(102.282, 65.6762 ){\s}\put(103.087, 65.5701
){\s}\put(103.893, 65.4679 ){\s}\put(104.698, 65.3694
){\s}\put(105.503, 65.2743 ){\s}\put(106.309, 65.1826
){\s}\put(107.114, 65.094 ){\s}\put(107.919, 65.0084
){\s}\put(108.725, 64.9256 ){\s}\put(109.53, 64.8455
){\s}\put(110.335, 64.768 ){\s}\put(111.141, 64.6929
){\s}\put(111.946, 64.6202 ){\s}\put(112.752, 64.5496
){\s}\put(113.557, 64.4812 ){\s}\put(114.362, 64.4148
){\s}\put(115.168, 64.3504 ){\s}\put(115.973, 64.2878
){\s}\put(116.778, 64.227 ){\s}\put(117.584, 64.1678
){\s}\put(118.389, 64.1104 ){\s}\put(120, 64){\s}


\put(0.0,  6.0){\s}\put(0.805369,   5.94558 ){\s}\put(1.61074,
5.88966 ){\s}\put(2.41611, 5.83217 ){\s}\put(3.22148, 5.77305
){\s}\put(4.02685, 5.71223 ){\s}\put(4.83222, 5.64963
){\s}\put(5.63758, 5.58519 ){\s}\put(6.44295, 5.5188
){\s}\put(7.24832, 5.45038 ){\s}\put(8.05369, 5.37984
){\s}\put(8.85906, 5.30709 ){\s}\put(9.66443, 5.232
){\s}\put(10.4698, 5.15447 ){\s}\put(11.2752, 5.07438
){\s}\put(12.0805, 4.9916 ){\s}\put(12.8859, 4.90598
){\s}\put(13.6913, 4.81739 ){\s}\put(14.4966, 4.72566
){\s}\put(15.302, 4.63063 ){\s}\put(16.1074, 4.53211
){\s}\put(16.9128, 4.42991 ){\s}\put(17.7181, 4.32381
){\s}\put(18.5235, 4.21359 ){\s}\put(19.3289, 4.09901
){\s}\put(20.1342, 3.9798 ){\s}\put(20.9396, 3.85567
){\s}\put(21.745, 3.72632 ){\s}\put(22.5503, 3.5914
){\s}\put(23.3557, 3.45055 ){\s}\put(24.1611, 3.30337
){\s}\put(24.9664, 3.14943 ){\s}\put(25.7718, 2.98824
){\s}\put(26.5772, 2.81928 ){\s}\put(27.3825, 2.64198
){\s}\put(28.1879, 2.4557 ){\s}\put(28.9933, 2.25974
){\s}\put(29.7986, 2.05334 ){\s}\put(30.604, 1.83562
){\s}\put(31.4094, 1.60564 ){\s}\put(32.2148, 1.36232
){\s}\put(33.0201, 1.10448 ){\s}\put(33.8255, 0.830772
){\s}\put(34.6309, 0.539685 ){\s}\put(35.4362, 0.22951){\s}


\put(37.8523, 0 ){\s} \put(41.0738,0){\s} \put(44.2953, 0 ){\s}
\put(47.5168, 0){\s} \put(50.7383,0){\s} \put(53.9598,0 ){\s}
\put(57.1812, 0){\s} \put(60.4027,0){\s} \put(63.6242,0){\s}
\put(66.8457, 0){\s} \put(70.0671,0 ){\s} \put(73.2886,0){\s}
\put(76.5101, 0){\s} \put(79.7315, 0){\s} \put(82.953,0){\s}


\put(84.5637,    0.229496 ){\s}\put(85.3691,    0.53967
){\s}\put(86.1745,    0.830757 ){\s}\put(86.9798, 1.10447
){\s}\put(87.7852, 1.36231 ){\s}\put(88.5906, 1.60562
){\s}\put(89.3959, 1.8356 ){\s}\put(90.2013, 2.05332
){\s}\put(91.0067, 2.25973 ){\s}\put(91.812, 2.45568
){\s}\put(92.6174, 2.64196 ){\s}\put(93.4228, 2.81926
){\s}\put(94.2281, 2.98822 ){\s}\put(95.0335, 3.14941
){\s}\put(95.8389, 3.30336 ){\s}\put(96.6442, 3.45054
){\s}\put(97.4496, 3.59139 ){\s}\put(98.255, 3.7263
){\s}\put(99.0603, 3.85566 ){\s}\put(99.8657, 3.97979
){\s}\put(100.671, 4.099 ){\s}\put(101.476, 4.21358
){\s}\put(102.282, 4.3238 ){\s}\put(103.087, 4.42989
){\s}\put(103.893, 4.5321 ){\s}\put(104.698, 4.63062
){\s}\put(105.503, 4.72565 ){\s}\put(106.309, 4.81738
){\s}\put(107.114, 4.90597 ){\s}\put(107.919, 4.99159
){\s}\put(108.725, 5.07437 ){\s}\put(109.53, 5.15446
){\s}\put(110.335, 5.23199 ){\s}\put(111.141, 5.30708
){\s}\put(111.946, 5.37983 ){\s}\put(112.752, 5.45037
){\s}\put(113.557, 5.51879 ){\s}\put(114.362, 5.58518
){\s}\put(115.168, 5.64962 ){\s}\put(115.973, 5.71222
){\s}\put(116.778, 5.77304 ){\s}\put(117.584, 5.83216
){\s}\put(118.389, 5.88965 ){\s}\put(119.194, 5.94557
){\s}\put(120, 5.99999){\s}

\end{picture}\hspace{60pt}
\begin{picture}(150,100)(0,-20)
\put (0,-10){\vector(1,0){140}} \put(0,-10){\vector(0,1){90}}
\multiput(-3,0)(0,10){8}{\line(1,0){3}}
\multiput(30,-13)(30,0){4}{\line(0,1){3}} \put(-15,
57.5){$\varsigma c^2$} \put(-15, 7.5){$p\,_0$}
\put(90,65){$\varsigma\langle u^\prime_1 u^\prime_1 \rangle $}
\put(105,23){$\langle p\,\rangle$} \put(135,-20){$x$}\put(120,
90){}\put(-10, -13.5){0}


\put(0,  56 ){\s}\put(0.805369,   55.9456 ){\s}\put(1.61074,
55.8897 ){\s}\put(2.41611,    55.8322 ){\s}\put(3.22148,    55.773
){\s}\put(4.02685,    55.7122 ){\s}\put(4.83222,    55.6496
){\s}\put(5.63758,    55.5852 ){\s}\put(6.44295,    55.5188
){\s}\put(7.24832,    55.4504 ){\s}\put(8.05369,    55.3798
){\s}\put(8.85906,    55.3071 ){\s}\put(9.66443,    55.232
){\s}\put(10.4698,    55.1545 ){\s}\put(11.2752,    55.0744
){\s}\put(12.0805,    54.9916 ){\s}\put(12.8859,    54.906
){\s}\put(13.6913,    54.8174 ){\s}\put(14.4966,    54.7257
){\s}\put(15.302,     54.6306 ){\s}\put(16.1074,    54.5321
){\s}\put(16.9128,    54.4299 ){\s}\put(17.7181,    54.3238
){\s}\put(18.5235,    54.2136 ){\s}\put(19.3289,    54.099
){\s}\put(20.1342,    53.9798 ){\s}\put(20.9396,    53.8557
){\s}\put(21.745,     53.7263 ){\s}\put(22.5503,    53.5914
){\s}\put(23.3557,    53.4506 ){\s}\put(24.1611,    53.3034
){\s}\put(24.9664,    53.1494 ){\s}\put(25.7718,    52.9882
){\s}\put(26.5772,    52.8193 ){\s}\put(27.3825,    52.642
){\s}\put(28.1879,    52.4557 ){\s}\put(28.9933,    52.2597
){\s}\put(29.7986,    52.0533 ){\s}\put(30.604,     51.8356
){\s}\put(31.4094,    51.6056 ){\s}\put(32.2148,    51.3623
){\s}\put(33.0201,    51.1045 ){\s}\put(33.8255,    50.8308
){\s}\put(34.6309,    50.5397 ){\s}\put(35.4362,    50.2295){\s}



\put(84.5637,    50.2295 ){\s}\put(85.3691,    50.5397
){\s}\put(86.1745,    50.8308 ){\s}\put(86.9798,    51.1045
){\s}\put(87.7852,    51.3623 ){\s}\put(88.5906,    51.6056
){\s}\put(89.3959,    51.8356 ){\s}\put(90.2013,    52.0533
){\s}\put(91.0067,    52.2597 ){\s}\put(91.812,     52.4557
){\s}\put(92.6174,    52.642 ){\s}\put(93.4228,    52.8193
){\s}\put(94.2281,    52.9882 ){\s}\put(95.0335,    53.1494
){\s}\put(95.8389,    53.3034 ){\s}\put(96.6442,    53.4505
){\s}\put(97.4496,    53.5914 ){\s}\put(98.255,     53.7263
){\s}\put(99.0603,    53.8557 ){\s}\put(99.8657,    53.9798
){\s}\put(100.671,    54.099 ){\s}\put(101.476,    54.2136
){\s}\put(102.282,    54.3238 ){\s}\put(103.087,    54.4299
){\s}\put(103.893,    54.5321 ){\s}\put(104.698,    54.6306
){\s}\put(105.503,    54.7257 ){\s}\put(106.309,    54.8174
){\s}\put(107.114,    54.906 ){\s}\put(107.919,    54.9916
){\s}\put(108.725,    55.0744 ){\s}\put(109.53,     55.1545
){\s}\put(110.335,    55.232 ){\s}\put(111.141,    55.3071
){\s}\put(111.946,    55.3798 ){\s}\put(112.752,    55.4504
){\s}\put(113.557,    55.5188 ){\s}\put(114.362,    55.5852
){\s}\put(115.168,    55.6496 ){\s}\put(115.973,    55.7122
){\s}\put(116.778,    55.773 ){\s}\put(117.584,    55.8322
){\s}\put(118.389,    55.8896 ){\s}\put(119.194,    55.9456
){\s}\put(120,    56){\s}


\put(0,  14 ){\s}\put(0.805369,   14.0544 ){\s}\put(1.61074,
14.1103 ){\s}\put(2.41611,    14.1678 ){\s}\put(3.22148,    14.227
){\s}\put(4.02685,    14.2878 ){\s}\put(4.83222,    14.3504
){\s}\put(5.63758,    14.4148 ){\s}\put(6.44295,    14.4812
){\s}\put(7.24832,    14.5496 ){\s}\put(8.05369,    14.6202
){\s}\put(8.85906,    14.6929 ){\s}\put(9.66443,    14.768
){\s}\put(10.4698,    14.8455 ){\s}\put(11.2752,    14.9256
){\s}\put(12.0805,    15.0084 ){\s}\put(12.8859,    15.094
){\s}\put(13.6913,    15.1826 ){\s}\put(14.4966,    15.2743
){\s}\put(15.302,     15.3694 ){\s}\put(16.1074,    15.4679
){\s}\put(16.9128,    15.5701 ){\s}\put(17.7181,    15.6762
){\s}\put(18.5235,    15.7864 ){\s}\put(19.3289,    15.901
){\s}\put(20.1342,    16.0202 ){\s}\put(20.9396,    16.1443
){\s}\put(21.745,     16.2737 ){\s}\put(22.5503,    16.4086
){\s}\put(23.3557,    16.5494 ){\s}\put(24.1611,    16.6966
){\s}\put(24.9664,    16.8506 ){\s}\put(25.7718,    17.0118
){\s}\put(26.5772,    17.1807 ){\s}\put(27.3825,    17.358
){\s}\put(28.1879,    17.5443 ){\s}\put(28.9933,    17.7403
){\s}\put(29.7986,    17.9467 ){\s}\put(30.604,     18.1644
){\s}\put(31.4094,    18.3944 ){\s}\put(32.2148,    18.6377
){\s}\put(33.0201,    18.8955 ){\s}\put(33.8255,    19.1692
){\s}\put(34.6309,    19.4603 ){\s}\put(35.4362,    19.7705){\s}

\put(37.8523, 20 ){\s} \put(41.0738,20){\s} \put(44.2953, 20 ){\s}
\put(47.5168, 20){\s} \put(50.7383,20){\s} \put(53.9598, 20 ){\s}
\put(57.1812, 20){\s} \put(60.4027,20){\s} \put(63.6242, 20){\s}
\put(66.8457, 20){\s} \put(70.0671, 20 ){\s} \put(73.2886,20){\s}
\put(76.5101, 20){\s} \put(79.7315, 20){\s} \put(82.953,20){\s}

\put(84.5637,    19.7705 ){\s}\put(85.3691,    19.4603
){\s}\put(86.1745,    19.1692){\s} \put(86.9798,    18.8955
){\s}\put(87.7852,    18.6377 ){\s}\put(88.5906,    18.3944
){\s}\put(89.3959,    18.1644 ){\s}\put(90.2013,    17.9467
){\s}\put(91.0067,    17.7403 ){\s}\put(91.812,     17.5443
){\s}\put(92.6174,    17.358 ){\s}\put(93.4228,    17.1807
){\s}\put(94.2281,    17.0118 ){\s}\put(95.0335,    16.8506
){\s}\put(95.8389,    16.6966 ){\s}\put(96.6442,    16.5495
){\s}\put(97.4496,    16.4086 ){\s}\put(98.255,     16.2737
){\s}\put(99.0603,    16.1443 ){\s}\put(99.8657,    16.0202
){\s}\put(100.671,    15.901 ){\s}\put(101.476,    15.7864
){\s}\put(102.282,    15.6762 ){\s}\put(103.087,    15.5701
){\s}\put(103.893,    15.4679 ){\s}\put(104.698,    15.3694
){\s}\put(105.503,    15.2743 ){\s}\put(106.309,    15.1826
){\s}\put(107.114,    15.094 ){\s}\put(107.919,    15.0084
){\s}\put(108.725,    14.9256 ){\s}\put(109.53,     14.8455
){\s}\put(110.335,    14.768 ){\s}\put(111.141,    14.6929
){\s}\put(111.946,    14.6202 ){\s}\put(112.752,    14.5496
){\s}\put(113.557,    14.4812 ){\s}\put(114.362,    14.4148
){\s}\put(115.168,    14.3504 ){\s}\put(115.973,    14.2878
){\s}\put(116.778,    14.227 ){\s}\put(117.584,    14.1678
){\s}\put(118.389,    14.1104 ){\s}\put(119.194,    14.0544
){\s}\put(120,    14){\s}
\end{picture}
\caption{\label{fig1} The proton (left) and antiproton (right).
Dots on the pressure graph indicate the vapor in the bubble.}
\end{figure}
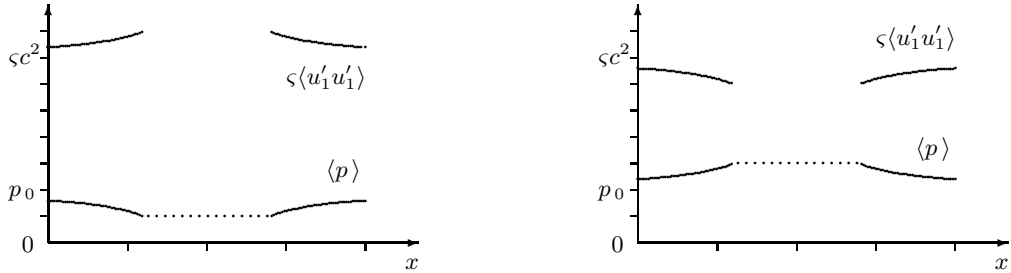

A turbulent fluid may adjust itself to boundary conditions via the
perturbation of the turbulence energy (Fig.\ref{fig1}, left). A
new stable configuration is thus formed. The conversion of the
neutron to the proton can be seen in the following way. A
microscopic restructuring takes place in the core of the particle.
The gas in the bubble cools down and the pressure on the wall
drops to some value $p_0-\Delta p$, where $\Delta p > 0$.

The field of the turbulence perturbation produced by the point
discontinuity was shown in \cite{Dmitriyev} to take the form
$1/r$. We have for the stable bubble of the radius $R_{\rm p}$
located at ${\bf x'}$:
\begin{eqnarray}
\left\langle p\, \right\rangle &=& p\,_0  - \frac{a}{{\left| {{\bf
x} - {\bf x'}} \right|}}, \label{32}\\
 a &=& R_{\rm p}\Delta p. \label{33}
\end{eqnarray}
The respective field of the turbulence energy can be found using
(\ref{32}) in (\ref{8}):
\begin{equation}
\varsigma\!\! \left\langle {u'_1 u'_1 } \right\rangle  = \varsigma
c^2 + \frac{a}{{\left| {{\bf x} - {\bf x'}} \right|}}. \label{34}
\end{equation}
The bubble inclusion that produces in the turbulent fluid the
field of the positive perturbation of the turbulence energy
models the proton (Fig.\ref{fig1}, left).

\section{The electron}

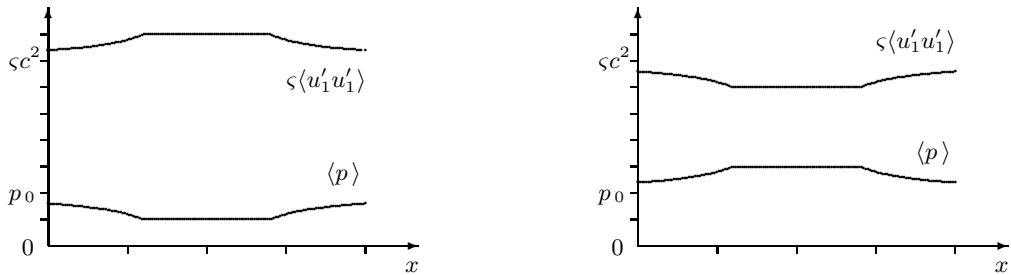
\begin{figure}[h]
\begin{picture}(160,100)(0,-20)
\put (0,-10){\vector(1,0){140}} \put(0,-10){\vector(0,1){90}}
\multiput(-3,0)(0,10){8}{\line(1,0){3}}
\multiput(30,-13)(30,0){4}{\line(0,1){3}}

\put(-15, 57.5){$\varsigma c^2$} \put(-15, 7.5){$p\,_0$}
\put(90,50){$\varsigma\langle u^\prime_1 u^\prime_1 \rangle $}
\put(105,15){$\langle p\,\rangle$} \put(135,-20){$x$} \put(120,
90){}\put(-10, -13.5){0}


\put(0.0,   64.0){\circle{0.01}}\put(0.805369,   64.0544
){\s}\put(1.61074, 64.1103 ){\s}\put(2.41611, 64.1678
){\s}\put(3.22148, 64.227 ){\s}\put(4.02685, 64.2878
){\s}\put(4.83222, 64.3504 ){\s}\put(5.63758, 64.4148
){\s}\put(6.44295, 64.4812 ){\s}\put(7.24832, 64.5496
){\s}\put(8.05369, 64.6202 ){\s}\put(8.85906, 64.6929
){\s}\put(9.66443, 64.768 ){\s}\put(10.4698, 64.8455
){\s}\put(11.2752, 64.9256 ){\s}\put(12.0805, 65.0084
){\s}\put(12.8859, 65.094 ){\s}\put(13.6913, 65.1826
){\s}\put(14.4966, 65.2743 ){\s}\put(15.302, 65.3694
){\s}\put(16.1074, 65.4679 ){\s}\put(16.9128, 65.5701
){\s}\put(17.7181, 65.6762 ){\s}\put(18.5235, 65.7864
){\s}\put(19.3289, 65.901 ){\s}\put(20.1342, 66.0202
){\s}\put(20.9396, 66.1443 ){\s}\put(21.745, 66.2737
){\s}\put(22.5503, 66.4086 ){\s}\put(23.3557, 66.5495
){\s}\put(24.1611, 66.6966 ){\s}\put(24.9664, 66.8506
){\s}\put(25.7718, 67.0118 ){\s}\put(26.5772, 67.1807
){\s}\put(27.3825, 67.358 ){\s}\put(28.1879, 67.5443
){\s}\put(28.9933, 67.7403 ){\s}\put(29.7986, 67.9467
){\s}\put(30.604, 68.1644 ){\s}\put(31.4094, 68.3944
){\s}\put(32.2148, 68.6377 ){\s}\put(33.0201, 68.8955
){\s}\put(33.8255, 69.1692 ){\s}\put(34.6309, 69.4603
){\s}\put(35.4362, 69.7705){\s}


\put(36.2416, 70 ){\s}\put(37.047, 70 ){\s}\put(37.8523, 70
){\s}\put(38.6577, 70){\s}\put(39.4631, 70 ){\s}\put(40.2685,
70){\s}\put(41.0738,70){\s}\put(41.8792, 70 ){\s}\put(42.6846,
70){\s}\put(43.4899,70){\s}\put(44.2953, 70 ){\s}\put(45.1007,
70){\s}\put(45.9061,70){\s}\put(46.7114, 70 ){\s}\put(47.5168,
70){\s}\put(48.3222,70){\s}\put(49.1275, 70 ){\s}\put(49.9329,
70){\s}\put(50.7383,70){\s}\put(51.5436, 70 ){\s}\put(52.349,
70){\s}\put(53.1544,70){\s}\put(53.9598, 70 ){\s}\put(54.7651,
70){\s}\put(55.5705,70){\s}


\put(56.3759,70){\s} \put(57.1812,70){\s} \put(57.9866,70){\s}
\put(58.792, 70){\s} \put(59.5973,  70){\s} \put(60.4027, 70){\s}
\put(61.2081, 70){\s} \put(62.0135, 70){\s} \put(62.8188, 70){\s}
\put(63.6242, 70){\s}


\put(64.4296, 70){\s}\put(65.2349, 70){\s}\put(66.0403,
70){\s}\put(66.8457,70){\s}\put(67.651,70){\s}\put(68.4564,
70){\s}\put(69.2618,70){\s}\put(70.0671,70){\s}\put(70.8725,
70){\s}\put(71.6779,70){\s}\put(72.4832,70){\s}\put(73.2886,
70){\s}\put(74.094, 70){\s}\put(74.8993,70){\s}\put(75.7047,
70){\s}\put(76.5101,70){\s}\put(77.3154,70){\s}\put(78.1208,
70){\s}\put(78.9262,70){\s}\put(79.7315,70){\s}\put(80.5369,
70){\s}\put(81.3423,70){\s}\put(82.1476,70){\s}\put(82.953,
70){\s}\put(83.7584,70){\s}

\put(84.5637,    69.7705 ){\s}\put(85.3691,    69.4603
){\s}\put(86.1745,    69.1692 ){\s}\put(86.9798, 68.8955
){\s}\put(87.7852, 68.6377 ){\s}\put(88.5906, 68.3944
){\s}\put(89.3959, 68.1644 ){\s}\put(90.2013, 67.9467
){\s}\put(91.0067, 67.7403 ){\s}\put(91.812, 67.5443
){\s}\put(92.6174, 67.358 ){\s}\put(93.4228, 67.1807
){\s}\put(94.2281, 67.0118 ){\s}\put(95.0335, 66.8506
){\s}\put(95.8389, 66.6966 ){\s}\put(96.6442, 66.5495
){\s}\put(97.4496, 66.4086 ){\s}\put(98.255, 66.2737
){\s}\put(99.0603, 66.1443 ){\s}\put(99.8657, 66.0202
){\s}\put(100.671, 65.901 ){\s}\put(101.476, 65.7864
){\s}\put(102.282, 65.6762 ){\s}\put(103.087, 65.5701
){\s}\put(103.893, 65.4679 ){\s}\put(104.698, 65.3694
){\s}\put(105.503, 65.2743 ){\s}\put(106.309, 65.1826
){\s}\put(107.114, 65.094 ){\s}\put(107.919, 65.0084
){\s}\put(108.725, 64.9256 ){\s}\put(109.53, 64.8455
){\s}\put(110.335, 64.768 ){\s}\put(111.141, 64.6929
){\s}\put(111.946, 64.6202 ){\s}\put(112.752, 64.5496
){\s}\put(113.557, 64.4812 ){\s}\put(114.362, 64.4148
){\s}\put(115.168, 64.3504 ){\s}\put(115.973, 64.2878
){\s}\put(116.778, 64.227 ){\s}\put(117.584, 64.1678
){\s}\put(118.389, 64.1104 ){\s}\put(120, 64){\s}


\put(0.0,  6.0){\s}\put(0.805369,   5.94558 ){\s}\put(1.61074,
5.88966 ){\s}\put(2.41611, 5.83217 ){\s}\put(3.22148, 5.77305
){\s}\put(4.02685, 5.71223 ){\s}\put(4.83222, 5.64963
){\s}\put(5.63758, 5.58519 ){\s}\put(6.44295, 5.5188
){\s}\put(7.24832, 5.45038 ){\s}\put(8.05369, 5.37984
){\s}\put(8.85906, 5.30709 ){\s}\put(9.66443, 5.232
){\s}\put(10.4698, 5.15447 ){\s}\put(11.2752, 5.07438
){\s}\put(12.0805, 4.9916 ){\s}\put(12.8859, 4.90598
){\s}\put(13.6913, 4.81739 ){\s}\put(14.4966, 4.72566
){\s}\put(15.302, 4.63063 ){\s}\put(16.1074, 4.53211
){\s}\put(16.9128, 4.42991 ){\s}\put(17.7181, 4.32381
){\s}\put(18.5235, 4.21359 ){\s}\put(19.3289, 4.09901
){\s}\put(20.1342, 3.9798 ){\s}\put(20.9396, 3.85567
){\s}\put(21.745, 3.72632 ){\s}\put(22.5503, 3.5914
){\s}\put(23.3557, 3.45055 ){\s}\put(24.1611, 3.30337
){\s}\put(24.9664, 3.14943 ){\s}\put(25.7718, 2.98824
){\s}\put(26.5772, 2.81928 ){\s}\put(27.3825, 2.64198
){\s}\put(28.1879, 2.4557 ){\s}\put(28.9933, 2.25974
){\s}\put(29.7986, 2.05334 ){\s}\put(30.604, 1.83562
){\s}\put(31.4094, 1.60564 ){\s}\put(32.2148, 1.36232
){\s}\put(33.0201, 1.10448 ){\s}\put(33.8255, 0.830772
){\s}\put(34.6309, 0.539685 ){\s}\put(35.4362, 0.22951){\s}



\put(36.2416, 0 ){\s}\put(37.047, 0 ){\s}\put(37.8523, 0
){\s}\put(38.6577, 0){\s}\put(39.4631, 0 ){\s}\put(40.2685,
0){\s}\put(41.0738,0){\s}\put(41.8792, 0 ){\s}\put(42.6846,
0){\s}\put(43.4899,0){\s}\put(44.2953, 0 ){\s}\put(45.1007,
0){\s}\put(45.9061,0){\s}\put(46.7114, 0 ){\s}\put(47.5168,
0){\s}\put(48.3222,0){\s}\put(49.1275, 0 ){\s}\put(49.9329,
0){\s}\put(50.7383,0){\s}\put(51.5436, 0 ){\s}\put(52.349,
0){\s}\put(53.1544,0){\s}\put(53.9598, 0 ){\s}\put(54.7651,
0){\s}\put(55.5705,0){\s}


\put(56.3759, 0){\s} \put(57.1812, 0){\s} \put(57.9866, 0){\s}
\put(58.792, 0){\s} \put(59.5973,  0){\s} \put(60.4027, 0){\s}
\put(61.2081, 0){\s} \put(62.0135, 0){\s} \put(62.8188, 0){\s}
\put(63.6242, 0){\s}


\put(64.4296, 0){\s}\put(65.2349, 0){\s}\put(66.0403,
0){\s}\put(66.8457,0){\s}\put(67.651,0 ){\s}\put(68.4564,
0){\s}\put(69.2618,0){\s}\put(70.0671,0 ){\s}\put(70.8725,
0){\s}\put(71.6779,0){\s}\put(72.4832,0 ){\s}\put(73.2886,
0){\s}\put(74.094, 0){\s}\put(74.8993,0 ){\s}\put(75.7047,
0){\s}\put(76.5101,0){\s}\put(77.3154,0 ){\s}\put(78.1208,
0){\s}\put(78.9262,0){\s}\put(79.7315,0 ){\s}\put(80.5369,
0){\s}\put(81.3423,0){\s}\put(82.1476,0 ){\s}\put(82.953,
0){\s}\put(83.7584,0){\s}


\put(84.5637,    0.229496 ){\s}\put(85.3691,    0.53967
){\s}\put(86.1745,    0.830757 ){\s}\put(86.9798, 1.10447
){\s}\put(87.7852, 1.36231 ){\s}\put(88.5906, 1.60562
){\s}\put(89.3959, 1.8356 ){\s}\put(90.2013, 2.05332
){\s}\put(91.0067, 2.25973 ){\s}\put(91.812, 2.45568
){\s}\put(92.6174, 2.64196 ){\s}\put(93.4228, 2.81926
){\s}\put(94.2281, 2.98822 ){\s}\put(95.0335, 3.14941
){\s}\put(95.8389, 3.30336 ){\s}\put(96.6442, 3.45054
){\s}\put(97.4496, 3.59139 ){\s}\put(98.255, 3.7263
){\s}\put(99.0603, 3.85566 ){\s}\put(99.8657, 3.97979
){\s}\put(100.671, 4.099 ){\s}\put(101.476, 4.21358
){\s}\put(102.282, 4.3238 ){\s}\put(103.087, 4.42989
){\s}\put(103.893, 4.5321 ){\s}\put(104.698, 4.63062
){\s}\put(105.503, 4.72565 ){\s}\put(106.309, 4.81738
){\s}\put(107.114, 4.90597 ){\s}\put(107.919, 4.99159
){\s}\put(108.725, 5.07437 ){\s}\put(109.53, 5.15446
){\s}\put(110.335, 5.23199 ){\s}\put(111.141, 5.30708
){\s}\put(111.946, 5.37983 ){\s}\put(112.752, 5.45037
){\s}\put(113.557, 5.51879 ){\s}\put(114.362, 5.58518
){\s}\put(115.168, 5.64962 ){\s}\put(115.973, 5.71222
){\s}\put(116.778, 5.77304 ){\s}\put(117.584, 5.83216
){\s}\put(118.389, 5.88965 ){\s}\put(119.194, 5.94557
){\s}\put(120, 5.99999){\s}

\end{picture}\hspace{60pt}
\begin{picture}(150,100)(0,-20)
\put (0,-10){\vector(1,0){140}} \put(0,-10){\vector(0,1){90}}
\multiput(-3,0)(0,10){8}{\line(1,0){3}}
\multiput(30,-13)(30,0){4}{\line(0,1){3}} \put(-15,
57.5){$\varsigma c^2$} \put(-15, 7.5){$p\,_0$}
\put(90,65){$\varsigma\langle u^\prime_1 u^\prime_1 \rangle $}
\put(105,23){$\langle p\,\rangle$} \put(135,-20){$x$}\put(120,
90){}\put(-10, -13.5){0}


\put(0,  56 ){\s}\put(0.805369,   55.9456 ){\s}\put(1.61074,
55.8897 ){\s}\put(2.41611,    55.8322 ){\s}\put(3.22148,    55.773
){\s}\put(4.02685,    55.7122 ){\s}\put(4.83222,    55.6496
){\s}\put(5.63758,    55.5852 ){\s}\put(6.44295,    55.5188
){\s}\put(7.24832,    55.4504 ){\s}\put(8.05369,    55.3798
){\s}\put(8.85906,    55.3071 ){\s}\put(9.66443,    55.232
){\s}\put(10.4698,    55.1545 ){\s}\put(11.2752,    55.0744
){\s}\put(12.0805,    54.9916 ){\s}\put(12.8859,    54.906
){\s}\put(13.6913,    54.8174 ){\s}\put(14.4966,    54.7257
){\s}\put(15.302,     54.6306 ){\s}\put(16.1074,    54.5321
){\s}\put(16.9128,    54.4299 ){\s}\put(17.7181,    54.3238
){\s}\put(18.5235,    54.2136 ){\s}\put(19.3289,    54.099
){\s}\put(20.1342,    53.9798 ){\s}\put(20.9396,    53.8557
){\s}\put(21.745,     53.7263 ){\s}\put(22.5503,    53.5914
){\s}\put(23.3557,    53.4506 ){\s}\put(24.1611,    53.3034
){\s}\put(24.9664,    53.1494 ){\s}\put(25.7718,    52.9882
){\s}\put(26.5772,    52.8193 ){\s}\put(27.3825,    52.642
){\s}\put(28.1879,    52.4557 ){\s}\put(28.9933,    52.2597
){\s}\put(29.7986,    52.0533 ){\s}\put(30.604,     51.8356
){\s}\put(31.4094,    51.6056 ){\s}\put(32.2148,    51.3623
){\s}\put(33.0201,    51.1045 ){\s}\put(33.8255,    50.8308
){\s}\put(34.6309,    50.5397 ){\s}\put(35.4362,    50.2295){\s}



\put(36.2416,    50 ){\s}\put(37.047,     50 ){\s}\put(37.8523, 50
){\s}\put(38.6577,    50){\s}\put(39.4631, 50 ){\s}\put(40.2685,
50){\s}\put(41.0738, 50){\s}\put(41.8792,  50 ){\s}\put(42.6846,
50){\s}\put(43.4899,  50){\s}\put(44.2953, 50 ){\s}\put(45.1007,
50){\s}\put(45.9061, 50){\s}\put(46.7114, 50 ){\s}\put(47.5168,
50){\s}\put(48.3222, 50){\s}\put(49.1275, 50 ){\s}\put(49.9329,
50){\s}\put(50.7383, 50){\s}\put(51.5436, 50 ){\s}\put(52.349,
50){\s}\put(53.1544, 50){\s}\put(53.9598, 50 ){\s}\put(54.7651,
50){\s}\put(55.5705, 50){\s}


\put(56.3759,  50){\s} \put(57.1812,  50){\s} \put(57.9866,
50){\s} \put(58.792, 50){\s} \put(59.5973,  50){\s} \put(60.4027,
50){\s} \put(61.2081, 50){\s} \put(62.0135, 50){\s} \put(62.8188,
50){\s} \put(63.6242, 50){\s}


\put(64.4296, 50){\s}\put(65.2349, 50){\s}\put(66.0403,
50){\s}\put(66.8457, 50){\s}\put(67.651, 50 ){\s}\put(68.4564,
50){\s}\put(69.2618, 50){\s}\put(70.0671, 50 ){\s}\put(70.8725,
50){\s}\put(71.6779, 50){\s}\put(72.4832, 50 ){\s}\put(73.2886,
50){\s}\put(74.094,  50){\s}\put(74.8993, 50 ){\s}\put(75.7047,
50){\s}\put(76.5101, 50){\s}\put(77.3154, 50 ){\s}\put(78.1208,
50){\s}\put(78.9262, 50){\s}\put(79.7315, 50 ){\s}\put(80.5369,
50){\s}\put(81.3423, 50){\s}\put(82.1476, 50 ){\s}\put(82.953,
50){\s}\put(83.7584,  50){\s}


\put(84.5637,    50.2295 ){\s}\put(85.3691,    50.5397
){\s}\put(86.1745,    50.8308 ){\s}\put(86.9798,    51.1045
){\s}\put(87.7852,    51.3623 ){\s}\put(88.5906,    51.6056
){\s}\put(89.3959,    51.8356 ){\s}\put(90.2013,    52.0533
){\s}\put(91.0067,    52.2597 ){\s}\put(91.812,     52.4557
){\s}\put(92.6174,    52.642 ){\s}\put(93.4228,    52.8193
){\s}\put(94.2281,    52.9882 ){\s}\put(95.0335,    53.1494
){\s}\put(95.8389,    53.3034 ){\s}\put(96.6442,    53.4505
){\s}\put(97.4496,    53.5914 ){\s}\put(98.255,     53.7263
){\s}\put(99.0603,    53.8557 ){\s}\put(99.8657,    53.9798
){\s}\put(100.671,    54.099 ){\s}\put(101.476,    54.2136
){\s}\put(102.282,    54.3238 ){\s}\put(103.087,    54.4299
){\s}\put(103.893,    54.5321 ){\s}\put(104.698,    54.6306
){\s}\put(105.503,    54.7257 ){\s}\put(106.309,    54.8174
){\s}\put(107.114,    54.906 ){\s}\put(107.919,    54.9916
){\s}\put(108.725,    55.0744 ){\s}\put(109.53,     55.1545
){\s}\put(110.335,    55.232 ){\s}\put(111.141,    55.3071
){\s}\put(111.946,    55.3798 ){\s}\put(112.752,    55.4504
){\s}\put(113.557,    55.5188 ){\s}\put(114.362,    55.5852
){\s}\put(115.168,    55.6496 ){\s}\put(115.973,    55.7122
){\s}\put(116.778,    55.773 ){\s}\put(117.584,    55.8322
){\s}\put(118.389,    55.8896 ){\s}\put(119.194,    55.9456
){\s}\put(120,    56){\s}


\put(0,  14 ){\s}\put(0.805369,   14.0544 ){\s}\put(1.61074,
14.1103 ){\s}\put(2.41611,    14.1678 ){\s}\put(3.22148,    14.227
){\s}\put(4.02685,    14.2878 ){\s}\put(4.83222,    14.3504
){\s}\put(5.63758,    14.4148 ){\s}\put(6.44295,    14.4812
){\s}\put(7.24832,    14.5496 ){\s}\put(8.05369,    14.6202
){\s}\put(8.85906,    14.6929 ){\s}\put(9.66443,    14.768
){\s}\put(10.4698,    14.8455 ){\s}\put(11.2752,    14.9256
){\s}\put(12.0805,    15.0084 ){\s}\put(12.8859,    15.094
){\s}\put(13.6913,    15.1826 ){\s}\put(14.4966,    15.2743
){\s}\put(15.302,     15.3694 ){\s}\put(16.1074,    15.4679
){\s}\put(16.9128,    15.5701 ){\s}\put(17.7181,    15.6762
){\s}\put(18.5235,    15.7864 ){\s}\put(19.3289,    15.901
){\s}\put(20.1342,    16.0202 ){\s}\put(20.9396,    16.1443
){\s}\put(21.745,     16.2737 ){\s}\put(22.5503,    16.4086
){\s}\put(23.3557,    16.5494 ){\s}\put(24.1611,    16.6966
){\s}\put(24.9664,    16.8506 ){\s}\put(25.7718,    17.0118
){\s}\put(26.5772,    17.1807 ){\s}\put(27.3825,    17.358
){\s}\put(28.1879,    17.5443 ){\s}\put(28.9933,    17.7403
){\s}\put(29.7986,    17.9467 ){\s}\put(30.604,     18.1644
){\s}\put(31.4094,    18.3944 ){\s}\put(32.2148,    18.6377
){\s}\put(33.0201,    18.8955 ){\s}\put(33.8255,    19.1692
){\s}\put(34.6309,    19.4603 ){\s}\put(35.4362,    19.7705){\s}


\put(36.2416, 20){\s}\put(37.047, 20){\s}\put(37.8523, 20
){\s}\put(38.6577, 20){\s}\put(39.4631, 20){\s}\put(40.2685,
20){\s}\put(41.0738, 20){\s}\put(41.8792,  20){\s}\put(42.6846,
20){\s}\put(43.4899, 20){\s}\put(44.2953, 20){\s}\put(45.1007,
20){\s}\put(45.9061, 20){\s}\put(46.7114,  20 ){\s}\put(47.5168,
20){\s}\put(48.3222, 20){\s}\put(49.1275, 20 ){\s}\put(49.9329,
20){\s}\put(50.7383, 20){\s}\put(51.5436, 20){\s}\put(52.349,
20){\s}\put(53.1544, 20){\s}\put(53.9598, 20 ){\s}\put(54.7651,
20){\s}\put(55.5705, 20){\s}


\put(56.3759, 20){\s} \put(57.1812, 20){\s} \put(57.9866, 20){\s}
\put(58.792, 20){\s} \put(59.5973, 20){\s} \put(60.4027, 20){\s}
\put(61.2081, 20){\s} \put(62.0135, 20){\s} \put(62.8188, 20){\s}
\put(63.6242, 20){\s}


\put(64.4296, 20){\s} \put(65.2349, 20){\s}\put(66.0403, 20
){\s}\put(66.8457,  20){\s}\put(67.651, 20){\s}\put(68.4564,
20){\s}\put(69.2618, 20){\s}\put(70.0671, 20 ){\s}\put(70.8725,
20){\s}\put(71.6779, 20){\s}\put(72.4832, 20 ){\s}\put(73.2886,
20){\s}\put(74.094, 20){\s}\put(74.8993, 20 ){\s}\put(75.7047,
20){\s}\put(76.5101, 20){\s}\put(77.3154, 20 ){\s}\put(78.1208,
20){\s}\put(78.9262, 20){\s}\put(79.7315, 20 ){\s}\put(80.5369,
20){\s}\put(81.3423, 20){\s}\put(82.1476, 20 ){\s}\put(82.953,
20){\s}\put(83.7584, 20){\s}


\put(84.5637,    19.7705 ){\s}\put(85.3691,    19.4603
){\s}\put(86.1745,    19.1692){\s} \put(86.9798,    18.8955
){\s}\put(87.7852,    18.6377 ){\s}\put(88.5906,    18.3944
){\s}\put(89.3959,    18.1644 ){\s}\put(90.2013,    17.9467
){\s}\put(91.0067,    17.7403 ){\s}\put(91.812,     17.5443
){\s}\put(92.6174,    17.358 ){\s}\put(93.4228,    17.1807
){\s}\put(94.2281,    17.0118 ){\s}\put(95.0335,    16.8506
){\s}\put(95.8389,    16.6966 ){\s}\put(96.6442,    16.5495
){\s}\put(97.4496,    16.4086 ){\s}\put(98.255,     16.2737
){\s}\put(99.0603,    16.1443 ){\s}\put(99.8657,    16.0202
){\s}\put(100.671,    15.901 ){\s}\put(101.476,    15.7864
){\s}\put(102.282,    15.6762 ){\s}\put(103.087,    15.5701
){\s}\put(103.893,    15.4679 ){\s}\put(104.698,    15.3694
){\s}\put(105.503,    15.2743 ){\s}\put(106.309,    15.1826
){\s}\put(107.114,    15.094 ){\s}\put(107.919,    15.0084
){\s}\put(108.725,    14.9256 ){\s}\put(109.53,     14.8455
){\s}\put(110.335,    14.768 ){\s}\put(111.141,    14.6929
){\s}\put(111.946,    14.6202 ){\s}\put(112.752,    14.5496
){\s}\put(113.557,    14.4812 ){\s}\put(114.362,    14.4148
){\s}\put(115.168,    14.3504 ){\s}\put(115.973,    14.2878
){\s}\put(116.778,    14.227 ){\s}\put(117.584,    14.1678
){\s}\put(118.389,    14.1104 ){\s}\put(119.194,    14.0544
){\s}\put(120,    14){\s}
\end{picture}
\caption{\label{fig2} The positron (left) and electron (right).}
\end{figure}

The form (\ref{34}) implies an infinite quantity of the total
energy perturbation
\begin{equation}
\frac{1}{2}\int\limits_\Omega {\varsigma \left( {\left\langle
{u'_i u'_i } \right\rangle  - \left\langle {u'_i u'_i }
\right\rangle ^{\left( 0 \right)} } \right)} d^3 x\label{35}
\end{equation}
where the medium volume $\Omega \to \infty$. So, the positive
deviation (\ref{34}) from the background (\ref{6}) should be
compensated by a negative deviation of the turbulence energy of
the same form, yet with the opposite sign of the coefficient
(\ref{33}):
\begin{equation}
a = - R_{\rm p}\Delta p. \label{36}
\end{equation}
Supposing that the energy attains a reduced value
$\varsigma\left\langle {u'_1 u'_1 } \right\rangle = \varsigma c^2
- \Delta\varepsilon$ at the core $\left| {{\bf x} - {\bf x'}}
\right| = r{}_{\rm e}$ of the negative disturbance center, we find
from (\ref{34}):
\begin{equation}
a =  -\, r_{\rm e} \Delta\varepsilon . \label{37}
\end{equation}
The center of the negative perturbation of the turbulence energy
serves as a model of the electron (Fig.\ref{fig2}, right).

\section{The antineutrino}

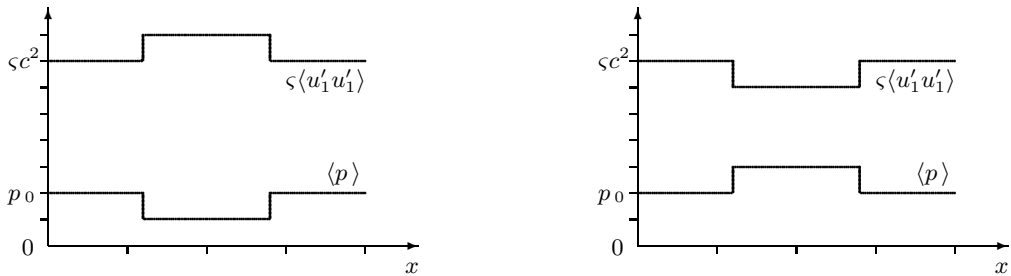
\begin{figure}[h]
\begin{picture}(160,100)(0,-20)
\put (0,-10){\vector(1,0){140}} \put(0,-10){\vector(0,1){90}}
\multiput(-3,0)(0,10){8}{\line(1,0){3}}
\multiput(30,-13)(30,0){4}{\line(0,1){3}}

\put(-15, 57.5){$\varsigma c^2$} \put(-15, 7.5){$p\,_0$}
\put(90,50){$\varsigma\langle u^\prime_1 u^\prime_1 \rangle $}
\put(105,15){$\langle p\,\rangle$} \put(135,-20){$x$} \put(120,
90){}\put(-10, -13.5){0}


\put(0,  60 ){\s}\put(0.805369,   60 ){\s}\put(1.61074,    60
){\s}\put(2.41611,    60 ){\s}\put(3.22148,    60
){\s}\put(4.02685, 60 ){\s}\put(4.83222,    60 ){\s}\put(5.63758,
60 ){\s}\put(6.44295,    60 ){\s}\put(7.24832,    60
){\s}\put(8.05369, 60 ){\s}\put(8.85906,    60 ){\s}\put(9.66443,
60 ){\s}\put(10.4698, 60 ){\s}\put(11.2752,    60
){\s}\put(12.0805, 60 ){\s}\put(12.8859, 60 ){\s}\put(13.6913, 60
){\s}\put(14.4966, 60 ){\s}\put(15.302, 60 ){\s}\put(16.1074, 60
){\s}\put(16.9128, 60 ){\s}\put(17.7181, 60 ){\s}\put(18.5235, 60
){\s}\put(19.3289, 60 ){\s}\put(20.1342, 60 ){\s}\put(20.9396, 60
){\s}\put(21.745, 60 ){\s}\put(22.5503,    60 ){\s}\put(23.3557,
60 ){\s}\put(24.1611, 60 ){\s}\put(24.9664,    60
){\s}\put(25.7718, 60 ){\s}\put(26.5772, 60 ){\s}\put(27.3825, 60
){\s}\put(28.1879, 60 ){\s}\put(28.9933, 60 ){\s}\put(29.7986, 60
){\s}\put(30.604, 60 ){\s}\put(31.4094, 60 ){\s}\put(32.2148, 60
){\s}\put(33.0201, 60 ){\s}\put(33.8255, 60 ){\s}\put(34.6309, 60
){\s}\put(35.4362, 60 ){\s}\put(35.5, 60.5){\s}

\put(36, 60.5 ){\s}\put(36, 61){\s}\put(36, 61.5 ){\s}\put(36,
62){\s}\put(36, 62.5){\s}\put(36, 63){\s}\put(36,
63.5){\s}\put(36, 64){\s}\put(36, 64.5){\s}\put(36,
65){\s}\put(36, 65.5){\s}\put(36, 66){\s}\put(36,
66.5){\s}\put(36, 67){\s}\put(36, 67.5){\s}\put(36,
68){\s}\put(36, 68.5){\s}\put(36, 69){\s}\put(36,
69.5){\s}\put(36, 70){\s}


\put(37.047, 70 ){\s}\put(37.8523, 70 ){\s}\put(38.6577,
70){\s}\put(39.4631, 70 ){\s}\put(40.2685,
70){\s}\put(41.0738,70){\s}\put(41.8792, 70 ){\s}\put(42.6846,
70){\s}\put(43.4899,70){\s}\put(44.2953, 70 ){\s}\put(45.1007,
70){\s}\put(45.9061,70){\s}\put(46.7114, 70 ){\s}\put(47.5168,
70){\s}\put(48.3222,70){\s}\put(49.1275, 70 ){\s}\put(49.9329,
70){\s}\put(50.7383,70){\s}\put(51.5436, 70 ){\s}\put(52.349,
70){\s}\put(53.1544,70){\s}\put(53.9598, 70 ){\s}\put(54.7651,
70){\s}\put(55.5705,70){\s}


\put(56.3759,70){\s} \put(57.1812,70){\s} \put(57.9866,70){\s}
\put(58.792, 70){\s} \put(59.5973,  70){\s} \put(60.4027, 70){\s}
\put(61.2081, 70){\s} \put(62.0135, 70){\s} \put(62.8188, 70){\s}
\put(63.6242, 70){\s}


\put(64.4296, 70){\s}\put(65.2349, 70){\s}\put(66.0403,
70){\s}\put(66.8457,70){\s}\put(67.651,70){\s}\put(68.4564,
70){\s}\put(69.2618,70){\s}\put(70.0671,70){\s}\put(70.8725,
70){\s}\put(71.6779,70){\s}\put(72.4832,70){\s}\put(73.2886,
70){\s}\put(74.094, 70){\s}\put(74.8993,70){\s}\put(75.7047,
70){\s}\put(76.5101,70){\s}\put(77.3154,70){\s}\put(78.1208,
70){\s}\put(78.9262,70){\s}\put(79.7315,70){\s}\put(80.5369,
70){\s}\put(81.3423,70){\s}\put(82.1476,70){\s}\put(82.953,
70){\s}\put(83.7584,70){\s}


\put(84, 60.5 ){\s}\put(84, 61){\s}\put(84, 61.5 ){\s}\put(84,
62){\s}\put(84, 62.5){\s}\put(84, 63){\s}\put(84,
63.5){\s}\put(84, 64){\s}\put(84, 64.5){\s}\put(84,
65){\s}\put(84, 65.5){\s}\put(84, 66){\s}\put(84, 66.5){\s}
\put(84,67){\s}\put(84, 67.5){\s}\put(84, 68){\s}\put(84,
68.5){\s}\put(84, 69){\s}\put(84, 69.5){\s}\put(84, 70){\s}

\put(84.5637, 60 ){\s}\put(85.3691,    60 ){\s}\put(86.1745, 60
){\s}\put(86.9798, 60 ){\s}\put(87.7852, 60 ){\s}\put(88.5906, 60
){\s}\put(89.3959, 60 ){\s}\put(90.2013, 60 ){\s}\put(91.0067, 60
){\s}\put(91.812, 60 ){\s}\put(92.6174, 60 ){\s}\put(93.4228, 60
){\s}\put(94.2281, 60 ){\s}\put(95.0335, 60 ){\s}\put(95.8389, 60
){\s}\put(96.6442, 60 ){\s}\put(97.4496, 60 ){\s}\put(98.255, 60
){\s}\put(99.0603, 60 ){\s}\put(99.8657, 60 ){\s}\put(100.671, 60
){\s}\put(101.476, 60 ){\s}\put(102.282, 60 ){\s}\put(103.087, 60
){\s}\put(103.893, 60 ){\s}\put(104.698, 60 ){\s}\put(105.503, 60
){\s}\put(106.309, 60 ){\s}\put(107.114, 60 ){\s}\put(107.919, 60
){\s}\put(108.725, 60 ){\s}\put(109.53, 60 ){\s}\put(110.335, 60
){\s}\put(111.141, 60 ){\s}\put(111.946, 60 ){\s}\put(112.752, 60
){\s}\put(113.557, 60 ){\s}\put(114.362, 60 ){\s}\put(115.168, 60
){\s}\put(115.973, 60 ){\s}\put(116.778, 60 ){\s}\put(117.584, 60
){\s}\put(118.389, 60 ){\s}\put(119.194, 60 ){\s}\put(120, 60){\s}


\put(0,  10 ){\s}\put(0.805369,   10 ){\s}\put(1.61074,    10
){\s}\put(2.41611,    10 ){\s}\put(3.22148,    10
){\s}\put(4.02685, 10 ){\s}\put(4.83222,    10 ){\s}\put(5.63758,
10 ){\s}\put(6.44295,    10 ){\s}\put(7.24832,    10
){\s}\put(8.05369, 10 ){\s}\put(8.85906,    10 ){\s}\put(9.66443,
10 ){\s}\put(10.4698, 10 ){\s}\put(11.2752,    10
){\s}\put(12.0805, 10 ){\s}\put(12.8859, 10 ){\s}\put(13.6913, 10
){\s}\put(14.4966, 10 ){\s}\put(15.302, 10 ){\s}\put(16.1074, 10
){\s}\put(16.9128, 10 ){\s}\put(17.7181, 10 ){\s}\put(18.5235, 10
){\s}\put(19.3289, 10 ){\s}\put(20.1342, 10 ){\s}\put(20.9396, 10
){\s}\put(21.745, 10 ){\s}\put(22.5503,    10 ){\s}\put(23.3557,
10 ){\s}\put(24.1611, 10 ){\s}\put(24.9664,    10
){\s}\put(25.7718, 10 ){\s}\put(26.5772, 10 ){\s}\put(27.3825, 10
){\s}\put(28.1879, 10 ){\s}\put(28.9933, 10 ){\s}\put(29.7986, 10
){\s}\put(30.604, 10 ){\s}\put(31.4094, 10 ){\s}\put(32.2148, 10
){\s}\put(33.0201, 10 ){\s}\put(33.8255, 10 ){\s}\put(34.6309, 10
){\s}\put(35.4362, 10 ){\s}

\put(36, 0.5 ){\s}\put(36, 1){\s}\put(36, 1.5 ){\s}\put(36,
2){\s}\put(36, 2.5){\s}\put(36, 3){\s}\put(36, 3.5){\s}\put(36,
4){\s}\put(36, 4.5){\s}\put(36, 5){\s}\put(36, 5.5){\s}\put(36,
6){\s}\put(36, 6.5){\s}\put(36, 7){\s}\put(36, 7.5){\s}\put(36,
8){\s}\put(36, 8.5){\s}\put(36, 9){\s}\put(36, 9.5){\s}\put(36,
10){\s}

\put(36.5,      00.0){\s}

\put(37.047, 0 ){\s}\put(37.8523, 0 ){\s}\put(38.6577,
0){\s}\put(39.4631,0){\s}\put(40.2685,
0){\s}\put(41.0738,0){\s}\put(41.8792, 0){\s}\put(42.6846,
0){\s}\put(43.4899,0){\s}\put(44.2953, 0){\s}\put(45.1007,
0){\s}\put(45.9061,0){\s}\put(46.7114, 0){\s}\put(47.5168,
0){\s}\put(48.3222,0){\s}\put(49.1275, 0){\s}\put(49.9329,
0){\s}\put(50.7383,0){\s}\put(51.5436, 0){\s}\put(52.349,
0){\s}\put(53.1544,0){\s}\put(53.9598, 0){\s}\put(54.7651,
0){\s}\put(55.5705,0){\s}


\put(56.3759,0){\s} \put(57.1812,0){\s} \put(57.9866,0){\s}
\put(58.792, 0){\s} \put(59.5973,0){\s} \put(60.4027,0){\s}
\put(61.2081, 0){\s} \put(62.0135,0){\s} \put(62.8188,0){\s}
\put(63.6242, 0){\s}


\put(64.4296,0){\s}\put(65.2349,0){\s}\put(66.0403,
0){\s}\put(66.8457,0){\s}\put(67.651,0){\s}\put(68.4564,
0){\s}\put(69.2618,0){\s}\put(70.0671,0){\s}\put(70.8725,
0){\s}\put(71.6779,0){\s}\put(72.4832,0){\s}\put(73.2886,
0){\s}\put(74.094, 0){\s}\put(74.8993,0){\s}\put(75.7047,
0){\s}\put(76.5101,0){\s}\put(77.3154,0){\s}\put(78.1208,
0){\s}\put(78.9262,0){\s}\put(79.7315,0){\s}\put(80.5369,
0){\s}\put(81.3423,0){\s}\put(82.1476,0){\s}\put(82.953,
0){\s}\put(83.7584,0){\s}


\put(84,0.5 ){\s}\put(84, 1){\s}\put(84, 1.5 ){\s}\put(84,
2){\s}\put(84, 2.5){\s}\put(84,3){\s}\put(84, 3.5){\s}\put(84,
4){\s}\put(84, 4.5){\s}\put(84, 5){\s}\put(84, 5.5){\s}\put(84,
6){\s}\put(84,6.5){\s} \put(84,7){\s}\put(84,7.5){\s}\put(84,
8){\s}\put(84,8.5){\s}\put(84,9){\s}\put(84,9.5){\s}\put(84,
10){\s}

\put(84.5637,    10 ){\s}\put(85.3691, 10 ){\s}\put(86.1745, 10
){\s}\put(86.9798, 10 ){\s}\put(87.7852, 10 ){\s}\put(88.5906, 10
){\s}\put(89.3959, 10 ){\s}\put(90.2013, 10 ){\s}\put(91.0067, 10
){\s}\put(91.812, 10 ){\s}\put(92.6174, 10 ){\s}\put(93.4228, 10
){\s}\put(94.2281, 10 ){\s}\put(95.0335, 10 ){\s}\put(95.8389, 10
){\s}\put(96.6442, 10 ){\s}\put(97.4496, 10 ){\s}\put(98.255, 10
){\s}\put(99.0603, 10 ){\s}\put(99.8657, 10 ){\s}\put(100.671, 10
){\s}\put(101.476, 10 ){\s}\put(102.282, 10 ){\s}\put(103.087, 10
){\s}\put(103.893, 10 ){\s}\put(104.698, 10 ){\s}\put(105.503, 10
){\s}\put(106.309, 10 ){\s}\put(107, 10 ){\s}\put(107.919, 10
){\s}\put(108.725, 10 ){\s}\put(109.53, 10 ){\s}\put(110.335, 10
){\s}\put(111.141, 10 ){\s}\put(111.946, 10 ){\s}\put(112.752, 10
){\s}\put(113.557, 10 ){\s}\put(114.362, 10 ){\s}\put(115.168, 10
){\s}\put(115.973, 10 ){\s}\put(116.778, 10 ){\s}\put(117.584, 10
){\s}\put(118.389, 10 ){\s}\put(119.194, 10 ){\s}\put(120, 10){\s}

\end{picture}\hspace{60pt}
\begin{picture}(160,100)(0,-20)
\put (0,-10){\vector(1,0){140}} \put(0,-10){\vector(0,1){90}}
\multiput(-3,0)(0,10){8}{\line(1,0){3}}
\multiput(30,-13)(30,0){4}{\line(0,1){3}}

\put(-15, 57.5){$\varsigma c^2$} \put(-15, 7.5){$p\,_0$}
\put(90,50){$\varsigma\langle u^\prime_1 u^\prime_1 \rangle $}
\put(105,15){$\langle p\,\rangle$} \put(135,-20){$x$} \put(120,
90){}\put(-10, -13.5){0}


\put(0,  60 ){\s}\put(0.805369,   60 ){\s}\put(1.61074,    60
){\s}\put(2.41611,    60 ){\s}\put(3.22148,    60
){\s}\put(4.02685, 60 ){\s}\put(4.83222,    60 ){\s}\put(5.63758,
60 ){\s}\put(6.44295,    60 ){\s}\put(7.24832,    60
){\s}\put(8.05369, 60 ){\s}\put(8.85906,    60 ){\s}\put(9.66443,
60 ){\s}\put(10.4698, 60 ){\s}\put(11.2752,    60
){\s}\put(12.0805, 60 ){\s}\put(12.8859, 60 ){\s}\put(13.6913, 60
){\s}\put(14.4966, 60 ){\s}\put(15.302, 60 ){\s}\put(16.1074, 60
){\s}\put(16.9128, 60 ){\s}\put(17.7181, 60 ){\s}\put(18.5235, 60
){\s}\put(19.3289, 60 ){\s}\put(20.1342, 60 ){\s}\put(20.9396, 60
){\s}\put(21.745, 60 ){\s}\put(22.5503,    60 ){\s}\put(23.3557,
60 ){\s}\put(24.1611, 60 ){\s}\put(24.9664,    60
){\s}\put(25.7718, 60 ){\s}\put(26.5772, 60 ){\s}\put(27.3825, 60
){\s}\put(28.1879, 60 ){\s}\put(28.9933, 60 ){\s}\put(29.7986, 60
){\s}\put(30.604, 60 ){\s}\put(31.4094, 60 ){\s}\put(32.2148, 60
){\s}\put(33.0201, 60 ){\s}\put(33.8255, 60 ){\s}\put(34.6309, 60
){\s}\put(35.4362, 60 ){\s}

\put(36, 50.5 ){\s}\put(36, 51){\s}\put(36, 51.5 ){\s}\put(36,
52){\s}\put(36, 52.5){\s}\put(36, 53){\s}\put(36,
53.5){\s}\put(36, 54){\s}\put(36, 54.5){\s}\put(36,
55){\s}\put(36, 55.5){\s}\put(36, 56){\s}\put(36,
56.5){\s}\put(36, 57){\s}\put(36, 57.5){\s}\put(36,
58){\s}\put(36, 58.5){\s}\put(36, 59){\s}\put(36,
59.5){\s}\put(36, 60){\s}

\put(36.0,      50.0){\s}

\put(37.047, 50 ){\s}\put(37.8523, 50 ){\s}\put(38.6577,
50){\s}\put(39.4631,50){\s}\put(40.2685,
50){\s}\put(41.0738,50){\s}\put(41.8792, 50){\s}\put(42.6846,
50){\s}\put(43.4899,50){\s}\put(44.2953, 50){\s}\put(45.1007,
50){\s}\put(45.9061,50){\s}\put(46.7114, 50){\s}\put(47.5168,
50){\s}\put(48.3222,50){\s}\put(49.1275, 50){\s}\put(49.9329,
50){\s}\put(50.7383,50){\s}\put(51.5436, 50){\s}\put(52.349,
50){\s}\put(53.1544,50){\s}\put(53.9598, 50){\s}\put(54.7651,
50){\s}\put(55.5705,50){\s}


\put(56.3759,50){\s} \put(57.1812,50){\s} \put(57.9866,50){\s}
\put(58.792, 50){\s} \put(59.5973, 50){\s} \put(60.4027, 50){\s}
\put(61.2081, 50){\s} \put(62.0135, 50){\s} \put(62.8188, 50){\s}
\put(63.6242, 50){\s}


\put(64.4296, 50){\s}\put(65.2349, 50){\s}\put(66.0403,
50){\s}\put(66.8457,50){\s}\put(67.651,50){\s}\put(68.4564,
50){\s}\put(69.2618,50){\s}\put(70.0671,50){\s}\put(70.8725,
50){\s}\put(71.6779,50){\s}\put(72.4832,50){\s}\put(73.2886,
50){\s}\put(74.094, 50){\s}\put(74.8993,50){\s}\put(75.7047,
50){\s}\put(76.5101,50){\s}\put(77.3154,50){\s}\put(78.1208,
50){\s}\put(78.9262,50){\s}\put(79.7315,50){\s}\put(80.5369,
50){\s}\put(81.3423,50){\s}\put(82.1476,50){\s}\put(82.953,
50){\s}\put(83.7584,50){\s}


\put(84, 50.5 ){\s}\put(84, 51){\s}\put(84, 51.5 ){\s}\put(84,
52){\s}\put(84, 52.5){\s}\put(84, 53){\s}\put(84,
53.5){\s}\put(84, 54){\s}\put(84, 54.5){\s}\put(84,
55){\s}\put(84, 55.5){\s}\put(84, 56){\s}\put(84, 56.5){\s}
\put(84,57){\s}\put(84, 57.5){\s}\put(84, 58){\s}\put(84,
58.5){\s}\put(84, 59){\s}\put(84, 59.5){\s}\put(84, 60){\s}

\put(84.5637,    60 ){\s}\put(85.3691, 60 ){\s}\put(86.1745, 60
){\s}\put(86.9798, 60 ){\s}\put(87.7852, 60 ){\s}\put(88.5906, 60
){\s}\put(89.3959, 60 ){\s}\put(90.2013, 60 ){\s}\put(91.0067, 60
){\s}\put(91.812, 60 ){\s}\put(92.6174, 60 ){\s}\put(93.4228, 60
){\s}\put(94.2281, 60 ){\s}\put(95.0335, 60 ){\s}\put(95.8389, 60
){\s}\put(96.6442, 60 ){\s}\put(97.4496, 60 ){\s}\put(98.255, 60
){\s}\put(99.0603, 60 ){\s}\put(99.8657, 60 ){\s}\put(100.671, 60
){\s}\put(101.476, 60 ){\s}\put(102.282, 60 ){\s}\put(103.087, 60
){\s}\put(103.893, 60 ){\s}\put(104.698, 60 ){\s}\put(105.503, 60
){\s}\put(106.309, 60 ){\s}\put(107.114, 60 ){\s}\put(107.919, 60
){\s}\put(108.725, 60 ){\s}\put(109.53, 60 ){\s}\put(110.335, 60
){\s}\put(111.141, 60 ){\s}\put(111.946, 60 ){\s}\put(112.752, 60
){\s}\put(113.557, 60 ){\s}\put(114.362, 60 ){\s}\put(115.168, 60
){\s}\put(115.973, 60 ){\s}\put(116.778, 60 ){\s}\put(117.584, 60
){\s}\put(118.389, 60 ){\s}\put(119.194, 60 ){\s}\put(120, 60){\s}


\put(0,  10 ){\s}\put(0.805369,   10 ){\s}\put(1.61074,    10
){\s}\put(2.41611,    10 ){\s}\put(3.22148,    10
){\s}\put(4.02685, 10 ){\s}\put(4.83222,    10 ){\s}\put(5.63758,
10 ){\s}\put(6.44295,    10 ){\s}\put(7.24832,    10
){\s}\put(8.05369, 10 ){\s}\put(8.85906,    10 ){\s}\put(9.66443,
10 ){\s}\put(10.4698, 10 ){\s}\put(11.2752,    10
){\s}\put(12.0805, 10 ){\s}\put(12.8859, 10 ){\s}\put(13.6913, 10
){\s}\put(14.4966, 10 ){\s}\put(15.302, 10 ){\s}\put(16.1074, 10
){\s}\put(16.9128, 10 ){\s}\put(17.7181, 10 ){\s}\put(18.5235, 10
){\s}\put(19.3289, 10 ){\s}\put(20.1342, 10 ){\s}\put(20.9396, 10
){\s}\put(21.745, 10 ){\s}\put(22.5503,    10 ){\s}\put(23.3557,
10 ){\s}\put(24.1611, 10 ){\s}\put(24.9664,    10
){\s}\put(25.7718, 10 ){\s}\put(26.5772, 10 ){\s}\put(27.3825, 10
){\s}\put(28.1879, 10 ){\s}\put(28.9933, 10 ){\s}\put(29.7986, 10
){\s}\put(30.604, 10 ){\s}\put(31.4094, 10 ){\s}\put(32.2148, 10
){\s}\put(33.0201, 10 ){\s}\put(33.8255, 10 ){\s}\put(34.6309, 10
){\s}\put(35.4362, 10 ){\s}

\put(36, 10.5 ){\s}\put(36, 11){\s}\put(36, 11.5 ){\s}\put(36,
12){\s}\put(36, 12.5){\s}\put(36, 13){\s}\put(36,
13.5){\s}\put(36, 14){\s}\put(36, 14.5){\s}\put(36,
15){\s}\put(36, 15.5){\s}\put(36, 16){\s}\put(36,
16.5){\s}\put(36, 17){\s}\put(36, 17.5){\s}\put(36,
18){\s}\put(36, 18.5){\s}\put(36, 19){\s}\put(36,
19.5){\s}\put(36, 20){\s}


\put(37.047, 20 ){\s}\put(37.8523, 20 ){\s}\put(38.6577,
20){\s}\put(39.4631,20){\s}\put(40.2685,
20){\s}\put(41.0738,20){\s}\put(41.8792, 20){\s}\put(42.6846,
20){\s}\put(43.4899,20){\s}\put(44.2953, 20){\s}\put(45.1007,
20){\s}\put(45.9061,20){\s}\put(46.7114, 20){\s}\put(47.5168,
20){\s}\put(48.3222,20){\s}\put(49.1275, 20){\s}\put(49.9329,
20){\s}\put(50.7383,20){\s}\put(51.5436, 20){\s}\put(52.349,
20){\s}\put(53.1544,20){\s}\put(53.9598, 20){\s}\put(54.7651,
20){\s}\put(55.5705,20){\s}


\put(56.3759,20){\s} \put(57.1812,20){\s} \put(57.9866,20){\s}
\put(58.792, 20){\s} \put(59.5973, 20){\s} \put(60.4027, 20){\s}
\put(61.2081, 20){\s} \put(62.0135, 20){\s} \put(62.8188, 20){\s}
\put(63.6242, 20){\s}


\put(64.4296, 20){\s}\put(65.2349, 20){\s}\put(66.0403,
20){\s}\put(66.8457,20){\s}\put(67.651,20){\s}\put(68.4564,
20){\s}\put(69.2618,20){\s}\put(70.0671,20){\s}\put(70.8725,
20){\s}\put(71.6779,20){\s}\put(72.4832,20){\s}\put(73.2886,
20){\s}\put(74.094, 20){\s}\put(74.8993,20){\s}\put(75.7047,
20){\s}\put(76.5101,20){\s}\put(77.3154,20){\s}\put(78.1208,
20){\s}\put(78.9262,20){\s}\put(79.7315,20){\s}\put(80.5369,
20){\s}\put(81.3423,20){\s}\put(82.1476,20){\s}\put(82.953,
20){\s}\put(83.7584,20){\s}


\put(84, 10.5 ){\s}\put(84, 11){\s}\put(84, 11.5 ){\s}\put(84,
12){\s}\put(84, 12.5){\s}\put(84, 13){\s}\put(84,
13.5){\s}\put(84, 14){\s}\put(84, 14.5){\s}\put(84,
15){\s}\put(84, 15.5){\s}\put(84, 16){\s}\put(84, 16.5){\s}
\put(84,17){\s}\put(84, 17.5){\s}\put(84, 18){\s}\put(84,
18.5){\s}\put(84, 19){\s}\put(84, 19.5){\s}\put(84, 20){\s}

\put(84.5637,    10 ){\s}\put(85.3691, 10 ){\s}\put(86.1745, 10
){\s}\put(86.9798, 10 ){\s}\put(87.7852, 10 ){\s}\put(88.5906, 10
){\s}\put(89.3959,    10 ){\s}\put(90.2013, 10 ){\s}\put(91.0067,
10 ){\s}\put(91.812, 10 ){\s}\put(92.6174, 10 ){\s}\put(93.4228,
10 ){\s}\put(94.2281, 10 ){\s}\put(95.0335, 10 ){\s}\put(95.8389,
10 ){\s}\put(96.6442, 10 ){\s}\put(97.4496, 10 ){\s}\put(98.255,
10 ){\s}\put(99.0603, 10 ){\s}\put(99.8657, 10 ){\s}\put(100.671,
10 ){\s}\put(101.476, 10 ){\s}\put(102.282, 10 ){\s}\put(103.087,
10 ){\s}\put(103.893, 10 ){\s}\put(104.698, 10 ){\s}\put(105.503,
10 ){\s}\put(106.309, 10 ){\s}\put(107.114, 10 ){\s}\put(107.919,
10 ){\s}\put(108.725, 10 ){\s}\put(109.53, 10 ){\s}\put(110.335,
10 ){\s}\put(111.141, 10 ){\s}\put(111.946, 10 ){\s}\put(112.752,
10 ){\s}\put(113.557, 10 ){\s}\put(114.362, 10 ){\s}\put(115.168,
10 ){\s}\put(115.973, 10 ){\s}\put(116.778, 10 ){\s}\put(117.584,
10 ){\s}\put(118.389, 10 ){\s}\put(119.194, 10 ){\s}\put(120,
10){\s}

\end{picture}
\caption{\label{fig3}The antineutrino (left) and neutrino
(right).}
\end{figure}

Comparing energy configurations of the proton (Fig.\ref{fig1},
left, top) and the electron (Fig.\ref{fig2}, right, top) we see
that the negative perturbation of the turbulence energy in the
core of the electron should be compensated by a respective
positive perturbation of the background energy (\ref{6}). This
surplus of the turbulence energy corresponds to the emission of a
new particle. Assuming in (\ref{36}) and (\ref{37}) $r_{\rm
e}\simeq R_{\rm p}$ we get $\Delta p \simeq \Delta \varepsilon$.
An isle of the fluid with increased turbulence energy may serve as
a model of the antineutrino (Fig.\ref{fig3}, left).

Thus, we constructed a linear mesoscopic mechanical scenario of
the reaction \begin{equation} n\, \to\, p^+ \,\,  +\, e^-
+\tilde{\nu}.\end{equation}

\section{Particles and waves}

Regions of the turbulent fluid such as in the models of the
neutrino and inner part of the electron are referred to in
hydrodynamics as the turbulence core. In the turbulence core the
simple closure $h_{ik}=0$ of the chain of governing equations
(\ref{1}), (\ref{2}) and (\ref{10}) is not valid. Using
appropriate closure schemes for (\ref{10}), perturbations of the
energy in the turbulence core was shown \cite{Troshkin1} to be in
general nonstationary. So, the neutrino as well as the electron
should be properly considered as waves of the turbulence energy.

The neutron and proton are associated with the inclusions into the
medium of the empty space. This is the reason why respective
perturbations of the turbulence energy are stationary, and the
baryon number is conserved.

On the other hand, in the model of the electromagnetic wave there
was shown \cite{Dmitriyev} to be no perturbation of the turbulence
energy. Nondiagonal terms of correlations $\left\langle {u'_i u'_k
} \right\rangle$ of turbulent fluctuations are only perturbed.

Thus we come to the conclusion that a particle is concerned with
the perturbation of the turbulence energy while the pure wave does
not accompanied by a perturbation of the turbulence energy.

\section{Quantum  mechanics}

There is considered a special kind of nonstationarity of the
wave-particles. The core of the electron is an isle of the fluid.
So, the electron can be additively decomposed into the collection
of fractions. Each fraction has an isle of the fluid in the core.
The size of the core is the same as in the whole particle. The
turbulence energy in the core is decreased relative to the
background level (\ref{6}). Yet it is greater than in the core of
the original localized electron. Each fraction generates in the
fluid the turbulence perturbation field that corresponds to a part
of the electric charge. The decomposition of the electron may
proceed under the action of turbulent fluctuations. This process
underlies what is known as quantum phenomena.

The proton can not be decomposed into fractions in the same
manner. It is a true particle.

\section{Antiparticles}

Models of antiparticles can be obtained mirroring graphs of
respective particles in relation to the  background levels of the
pressure (\ref{5}) and energy (\ref{6}). Thus we come to a model
of the positron as the center of the positive perturbation of the
turbulence energy (Fig.\ref{fig2}, left). The neutrino is the isle
of the fluid with the reduced turbulence energy (Fig.\ref{fig3},
right).

In the present approach the proton was modelled by the bubble with
vapor in it cooled as compared with the vapor in the neutron core.
Then the bubble with the vapor heated in comparison with the vapor
in the neutron core models the antiproton (Fig.\ref{fig1}, right).

\section{Conclusion}

In the mesoscopic approach the luminiferous medium can be modelled
by the averaged turbulence of an ideal fluid. Particles of matter
are concerned with inclusions into the medium of empty space. The
neutron is modelled by the vapor bubble. The cooling of the vapor
in the bubble forms the center of the positive perturbation of the
turbulence energy that serves as a model of the proton. The center
of the negative perturbation of the turbulence energy models the
electron. The antineutrino is concerned with a local positive
perturbation of the turbulence energy necessary in order to
compensate the difference in perturbations of the turbulence
energy between the proton and the electron.

\end{document}